\documentclass[prl,twocolumn,aps,showpacs,draft]{revtex4}
\newcommand{\bfm}[1]{\mbox{\boldmath${#1}$}}
\begin{document}
\title{Nonlinear Transformation for a Class of Gauged Schr\"odinger
Equations with Complex Nonlinearities}
\author{G. Kaniadakis}\email{kaniadakis@polito.it}
\author{A.M. Scarfone}\email{scarfone@polito.it}
\affiliation{Dipartimento di Fisica - Politecnico di Torino \\
Corso Duca degli Abruzzi 24, 10129 Torino, Italy; \\ Istituto
Nazionale di Fisica della Materia - Unit\'a del Politecnico di
Torino}
\date{\today}
\begin {abstract}
In the present contribution we consider a class of Schr\"odinger
equations containing complex nonlinearities, describing systems
with conserved norm $|\psi|^2$ and minimally coupled to an abelian
gauge field. We introduce a nonlinear transformation which permits
the linearization of the source term in the evolution equations
for the gauge field, and transforms the nonlinear Schr\"odinger
equations in another one with real nonlinearities. We show that
this transformation can be performed either on the gauge field
$A_\mu$ or, equivalently, on the matter field $\psi$. Since the
transformation does not change the quantities $|\psi|^2$ and
$F_{\mu\nu}$, it can be considered a generalization of the gauge
transformation of third kind introduced some years ago by other
authors.\\

\end {abstract}
\maketitle
In a previous paper \cite{Kaniadakis} we have
introduced a nonlinear unitary transformation acting on a class
of nonlinear Schr\"odinger equations (NLSEs) with complex
nonlinearities in order to transform these equations in other
NLSEs containing purely real nonlinearities. As a consequence,
the continuity equation for the new matter field results to be
linearized. This nonlinear transformation can be seen as a
generalization of the Doebner-Goldin approach adopted to obtain
the NLSE associated with unitary group representations
\cite{Doebner1,Doebner2,Doebner3,Nettermann}. The present paper is
a natural continuation of Ref. \cite{Kaniadakis}. We consider a
class of NLSEs minimally coupled with an abelian gauge field
$A_\mu$. The $(n+1)$ model is described by the following
Lagrangian density:
\begin{eqnarray}
{\cal L}={\cal L}_{\rm matter}+{\cal L}_{\rm gauge} \
,\label{lagrangiana}
\end{eqnarray}
with
\begin{eqnarray}
\nonumber {\cal L}_{\rm
matter}&=&i\,c\,\frac{\hbar}{2}\left[\psi^\ast\,D_0\psi-\psi\,(D_0\psi)^\ast\right]-
\frac{\hbar^2}{2\,m}\left|{\bfm
D}\psi\right|^2\\
&-&U([\psi],\,[\psi^\ast],\,{\bfm A}) \ ,
\end{eqnarray}
being $D_\mu=\partial_\mu+(i\,e/\hbar\,c)\,A_\mu\equiv(D_0,\,{\bfm
D})$ the covariant derivative with
$\partial_\mu\equiv(c^{-1}\,\partial_t,\,{\bfm\nabla})$,
$\mu=0,\,1,\,\cdots,\,n$.

The nonlinear potential $U([\psi],\,[\psi^\ast],\,{\bfm A})$ is
assumed to be real. We use the notation $U([a])$ to indicate that
$U$ is a function of the field $a$ and of its spatial covariant
derivatives. After introducing the hydrodynamic fields
$\rho(t,\,{\bfm x})$ and $S(t,\,{\bfm x})$ through:
\begin{eqnarray}
&&\rho=\psi\,\psi^\ast \ ,\\
&&S=i\,\frac{\hbar\,c}{2\,e}\,\log\left(\frac{\psi^\ast}{\psi}\right)
\ ,
\end{eqnarray}
which represent the modulo and the phase of the field $\psi$
respectively:
\begin{eqnarray}
\psi(t,\,{\bfm x})=\rho^{1/2}(t,\,{\bfm
x})\,\exp\left[\frac{i\,e}{\hbar\,c}\,S(t,\,{\bfm x})\right] \ ,
\end{eqnarray}
we can write the nonlinear potential as $U=U([\rho], \,[S],\,{\bfm
A})$.

In the following, for simplicity, we assume for the Lagrangian
${\cal L}_{\rm gauge}$ the standard form of the electromagnetic
field:
\begin{eqnarray}
{\cal L}_{\rm gauge}=-\frac{1}{4}\,F^{\mu\nu}\,F_{\mu\nu} \ ,
\end{eqnarray}
where the electromagnetic tensor is defined as
$F_{\mu\nu}=\partial_\mu\,A_\nu-\partial_\nu\,A_\mu$. We obtain
upper and lower indices by using the metric tensor
$\eta^{\mu\nu}=diag(1,\,-1,\,\cdots,\,-1)$:
$A^\mu=\eta^{\mu\nu}\,A_\nu$. Greek indices take the values
$0,\,1,\,2,\,\cdots,\,n$ while the latin indices take the values
$1,\,2,\,\cdots,\,n$.

Starting from the action of the system ${\cal A}=\int{\cal
L}\,dt\,d^nx$, the evolution equations for the fields
$a\equiv\psi,\,\psi^\ast,\,A_\mu,\,\rho$ and $S$ can be obtained
from the relation $\delta\,{\cal A}/\delta\,a=0$, where the
functional derivative is defined as \cite{Olver}:
\begin{eqnarray}
\frac{\delta}{\delta\,a}\,\int
F[a]\,dt\,d^nx=\sum_{k=0}(-1)^k\sum_I \,{\cal
D}_I\,\frac{\partial\,F[a]}{\partial\,({\cal D}_Ia)} \
.\label{der}
\end{eqnarray}
In Eq. (\ref{der}), the second sum is over the multi-index
$I\equiv(i_1,\,i_2,\,\cdots,\,i_n)$, and $0\leq i_p\leq k$, $\sum
i_p=k$, ${\cal D}_{_I}\equiv\partial^k/(\partial\,x^{i_1}_1\,
\cdots\,\partial\,x^{i_n}_{_n})$. When $a=\psi^\ast$ we obtain the
following gauged NLSE:
\begin{eqnarray}
i\,c\,\hbar\,D_0\,\psi=-\frac{\hbar^2}{2\,m}\,{\bfm
D}^2\,\psi+W\,\psi+i\,{\cal W}\,\psi \ ,\label{schroedinger}
\end{eqnarray}
where the real $W$ and imaginary ${\cal W}$ parts of the
nonlinearity in Eq. (\ref{schroedinger}) are given by:
\begin{eqnarray}
&&W([\rho],\,[S],\,{\bfm A})=\frac{\delta}{\delta\,\rho}\,\int
U([\rho],\,[S],\,{\bfm A})\,dt\,d^nx \ ,\\
\nonumber&&{\cal W}([\rho],\,[S],\,{\bfm
A})=\frac{\hbar\,c}{2\,e\,\rho}\,\frac{\delta}{\delta\,S}\,\int
U([\rho],\,[S],\,{\bfm A})\,dt\,d^nx \ .\\
\end{eqnarray}

Differently, for $a=S$ we obtain:
\begin{eqnarray}
\nonumber
&&\frac{\partial\,\rho}{\partial\,t}+{\bfm\nabla}\cdot\left[\frac{e}{m\,c}\,({\bfm\nabla}\,S-{\bfm
A})\,\rho\right]\\
&&-\frac{c}{e}\,\frac{\delta}{\delta\,S}
\,\int\,U([\rho],\,[S],\,{\bfm A})\,dt\,d^nx=0 \ ,
\end{eqnarray}
which can be rewrite in:
\begin{eqnarray}
\frac{\partial\,\rho}{\partial\,t}+{\bfm\nabla}\cdot{\bfm
J}=\frac{c}{e}\,\int\frac{\partial}{\partial\,S}\,U([\rho],\,[S],\,{\bfm
A})\,dt\,d^nx \ ,\label{eee}
\end{eqnarray}
where the current is given by
\begin{eqnarray}
J_i=\frac{e}{m\,c}\,(\partial_i\,S+A_i)\,\rho+\frac{c}{e}\,\frac{\delta}{\delta\,(\partial_i\,S)}
\,\int U\,\,dt\,d^nx .\label{current}
\end{eqnarray}
If the nonlinear potential $U$ depends on $S$ only trough its
derivatives, the r.h.s of Eq. (\ref{eee}) vanish and its became a
continuity equation of the field $\rho$:
\begin{eqnarray}
\frac{\partial\,\rho}{\partial\,t}+{\bfm\nabla}\cdot{\bfm J}=0 \
,\label{continuity}
\end{eqnarray}
which implies the conservation of the total charge:
\begin{eqnarray}
Q=e\int\rho\,d^nx \ .
\end{eqnarray}
Following the Noether theorem, Eq. (\ref{continuity}) implies the
existence of a symmetry for the system. The action $\cal A$ is
invariant over global transformation of the $U(1)$ group:
$\psi\rightarrow\psi^\prime=\psi\,\exp(i\,\epsilon)$. We remark
that this symmetry exists only if the nonlinear potential $U$
depend on $S$ only through its derivatives.

Analogously the evolution equation $\delta\,{\cal
A}/\delta\,A_\nu=0$ of the field $A_\nu$ becomes:
\begin{eqnarray}
\partial^\mu\,F_{\mu\nu}=\frac{e}{c}\,J_\nu \ ,\label{gauge}
\end{eqnarray}
where the covariant current $J_\nu$ is given by:
$J_\nu\equiv(c\,\rho,\,-{\bfm J})$.

Now we will introduce a nonlinear and nonlocal transformation
whose effect is to eliminate the imaginary part of the
nonlinearity in the NLSE so that the particle current associated
to the new transformed Schr\"odinger equation assumes the standard
bilinear structure. As it will be shown, this can be performed in
two different ways: or by a unitary transformation acting on the
field $\psi$ or by a gauge transformation acting on the field
$A_\mu$.

We start by considering the transformation acting on the field
$\psi$:
\begin{eqnarray}
\psi(t,\,{\bfm x})\rightarrow\phi(t,\,{\bfm x})={\cal
U}([\rho],\,[S],\,{\bfm A})\,\psi(t,\,{\bfm x}) \ ,\label{trasf1}
\end{eqnarray}
which allows to eliminate the imaginary part $\cal W$ of the
nonlinearity in the evolution equation (\ref{schroedinger}), and
to standardize the expression of the current:
\begin{eqnarray}
{\bfm J}([\rho],\,[S],\,{\bfm A})\rightarrow{\bfm{\cal J}}(
\rho,\,{\bfm\nabla}s,\,{\bfm A})=\frac{e}{m\,c}\,({\bfm
\nabla}s-{\bfm A})\,\rho \ .
\end{eqnarray}
The functional ${\cal U}$, with ${\cal U}^\ast={\cal U}^{-1}$, can
be written as:
\begin{eqnarray}
{\cal U}([\rho],\,[S],\,{\bfm
A})=\exp\left[\frac{i\,e}{\hbar\,c}\,\sigma\left([\rho],\,[S],\,{\bfm
A}\right)\right]
 \ ,\label{trasf2}
\end{eqnarray}
where the generator $\sigma$ of the transformation is defined
through the relation:
\begin{eqnarray}
\partial_i\sigma\left([\rho],\,[S],\,{\bfm
A}\right)=\frac{m\,c^2}{e^2\,\rho}\,\frac{\delta}{\delta\,(\partial_i\,S)}\int
U\,dt\,d^nx \ .\label{sig}
\end{eqnarray}
Eq. (\ref{sig}) imposes a condition on the form of the nonlinear
potential $U$which can be obtained using the relation:
$\partial_{ij}\,\sigma=\partial_{ji}\,\sigma$:
\begin{eqnarray}
\nonumber \left\{
\partial_i\left[{1\over\rho}\,\frac{\delta\,}{\delta\,(\partial_j\,S)}\right]-
\partial_j\left[{1\over\rho}\,\frac{\delta\,}{\delta\,(\partial_i\,S)}\right]\right\}\!\int
U\,dt\,d^nx=0 \ ,\label{condition}\\
\end{eqnarray}
for all $i,\,j=1,\,\cdots,\,n$. We remark that the condition
(\ref{condition}) select the potential $U([\rho],\,[S],\,{\bfm
A})$ and the nonlinear systems where we can perform the
transformation (\ref{trasf1}). From Eqs. (\ref{trasf1}) and
(\ref{trasf2}) it is easy to obtain the phase $s$ of the new field
$\phi$:
\begin{eqnarray}
s=S+\sigma([\rho],\,[S],\,{\bfm A}) \ ,\label{ns}
\end{eqnarray}
while the modulo of $\phi$ is equal to the modulo of $\psi$,
because of the unitariety of the transformation. If Eq. (\ref{ns})
is invertible, we can express the old phase $S$ as a functional of
$\rho,\,s$ and $\bfm A$: $S=S([\rho],\,[s],\,{\bfm A})$.\\ From
Eq. (\ref{schroedinger}) and taking into account Eqs.
(\ref{trasf1}), (\ref{trasf2}) and (\ref{sig}), it is easy to
obtain the following NLSE for the new field $\phi$:
\begin{eqnarray}
i\,c\,\hbar\,D_0\,\phi=-\frac{\hbar^2}{2\,m}\,{\bfm
D}^2\,\phi+\widetilde{W}([\rho],\,[s],\,{\bfm A})\,\phi \
,\label{schroedinger2}
\end{eqnarray}
where the real nonlinearity $\widetilde W$ is given by:
\begin{eqnarray}
\nonumber \widetilde{W}([\rho],\,[s],\,{\bfm
A})&=&W+\frac{e^2}{2\,m\,c^2}\,({\bfm\nabla}\,\sigma)^2\\
&-&\frac{e}{c}\,\frac{{\bfm{\cal
J}}\cdot{\bfm\nabla}\,\sigma}{\rho}-\frac{e}{c}\,\frac{\partial\,\sigma}{\partial\,t}
\ ,
\end{eqnarray}
with $W\equiv W([\rho],\,[S([\rho],\,[s],\,{\bfm A})$. As a
consequence of Eq. (\ref{schroedinger2}), the continuity equation
becomes:
\begin{eqnarray}
\frac{\partial\,\rho}{\partial\,t}+{\bfm\nabla}\cdot{\bfm{\cal
J}}=0 \ .
\end{eqnarray}
Finally, the evolution equation (\ref{gauge}) for the gauge field
$A_\mu$ becomes:
\begin{eqnarray}
\partial^\mu\,F_{\mu\nu}=\frac{e}{c}\,{\cal J}_\nu \ ,
\end{eqnarray}
where now also the source ${\cal
J}_\nu\equiv(c\,\rho,\,-{\bfm{\cal J}})$ results linearized.

We consider now the transformation acting on the gauge field
$A_\mu$:
\begin{eqnarray}
{\bfm A}\rightarrow{\bfm\chi}={\bfm A}-{\bfm\nabla}\sigma \ .
\label{trasfgauge1}
\end{eqnarray}
It is well know that the gauge field is the fundamental quantity
of the theory, while the physical informations are carried out by
the tensor field $F_{\mu\nu}$ \cite {Jackson}. We require that the
transformation leaves unchanged this quantity:
\begin{eqnarray}
F_{\mu\nu}\equiv\partial_\mu\,A_\nu-\partial_\nu\,A_\mu=
\partial_\mu\,\chi_\nu-\partial_\nu\,\chi_\mu \
.\label{condition2}
\end{eqnarray}
When $\mu,\,\nu$ are spatial indices, Eq. (\ref{condition2}) is
satisfied if Eq. (\ref{condition}) still holds. When $\mu$ or
$\nu$ is equal to zero, Eq. (\ref{condition2}) implies the
following transformation for $A_0$:
\begin{eqnarray}
A_0\rightarrow\chi_0=A_0+{1\over
c}\,\frac{\partial\,\sigma}{\partial\,t} \ , \label{trasfgauge2}
\end{eqnarray}
where $\sigma$ is given by Eq. (\ref{sig}). Then, from Eq.
(\ref{schroedinger}), and taking into account Eqs.
(\ref{trasfgauge1}) and (\ref{trasfgauge2}) we obtain:
\begin{eqnarray}
i\,c\,\hbar\,{\overline
D}_0\,\psi=-\frac{\hbar^2}{2\,m}\,\overline{\bfm
D}^{\,2}\,\psi+\overline{W}([\rho],\,[S],\,{\bfm \chi})\,\psi \ ,
\end{eqnarray}
which has the same form of Eq. (\ref{schroedinger2}) but
$\overline{D}_\mu=\partial_\mu+(i\,e/\hbar\,c)\,\chi_\mu$ and the
real nonlinearity is:
\begin{eqnarray}
\nonumber
\overline{W}([\rho],\,[S],\,{\bfm\chi})&=&W+\frac{e^2}{2\,m\,c^2}\,({\bfm\nabla}\sigma)^2\\
&-&\frac{e}{c}\,\frac{{{\bfm{\cal I
}}}\cdot{\bfm\nabla}\sigma}{\rho}-\frac{e}{c}\,\frac{\partial\,\sigma}{\partial\,t}
\ ,
\end{eqnarray}
with $W\equiv W([\rho],\,[S],\,{\bfm
\chi}+{\bfm\nabla}\,\sigma)],\,)$ and the new linearized current
\begin{eqnarray}
{\bfm{\cal I}}=\frac{e}{m\,c}\,({\bfm\nabla}\,S-{\bfm\chi})\,\rho
\ ,
\end{eqnarray}
obeys the continuity equation:
\begin{eqnarray}
\frac{\partial\,\rho}{\partial\,t}+{\bfm\nabla}\cdot{\bfm{\cal
I}}=0 \ .
\end{eqnarray}
On the same foot, the evolution equation (\ref{gauge}) for the
gauge field becomes:
\begin{eqnarray}
\partial^\mu\,F_{\mu\nu}=\frac{e}{c}\,{\cal I}_\nu \ ,
\end{eqnarray}
where ${\cal I}_\nu\equiv(c\,\rho,\,-{\bfm{\cal I}})$.

In order to show how the method above described can be used, we
consider the canonical subclass of the Doebner and Goldin
equations \cite{Doebner3,Nettermann}:
\begin{eqnarray}\nonumber
&&i\,\hbar\,\frac{\partial\,\psi}{\partial\,t}=-\frac{\hbar^2}{2\,m}\,\Delta\,\psi+
m\,\nu\,{\bfm\nabla}\left(\frac{{\bfm j}_0}{\rho}\right)\,\psi\\
&&-2\alpha\frac{\hbar^2}{m}\left[\frac{\Delta\,\rho}{\rho}
-{1\over2}\left(\frac{{\bfm\nabla}\,\rho}{\rho}\right)^2\right]\psi+i\frac{\hbar}{2}
\nu\frac{\Delta\,\rho}{\rho}\psi \ ,\label{dob}
\end{eqnarray}
where ${\bfm
j}_0=(-i\,\hbar/2\,m)\,(\psi^\ast\,{\bfm\nabla}\,\psi-\psi\,{\bfm\nabla}\,\psi^\ast)$
is the standard quantum-mechanical current, $\nu$ is a diffusion
coefficient, and $\alpha$ a dimensionless coupling constant. Eq.
(\ref{dob}) is obtainable starting from the nonlinear potential:
\begin{eqnarray}
U_{_{\rm DG}}([\rho],[S])=\frac{\nu}{2}\left(\rho\Delta
S-\!{\bfm\nabla}\rho\cdot\!
{\bfm\nabla}S\right)+\alpha\frac{\hbar^2}{m}\frac{({\bfm\nabla}\rho)^2}{\rho}\
\ .
\end{eqnarray}
It is easy to verify that the current ${\bfm j}$ associated to Eq.
(\ref{dob}) is given by:
\begin{eqnarray}
{\bfm j}=\frac{{\bfm\nabla}S}{m}\,\rho-\nu\,{\bfm\nabla}\rho \ ,
\end{eqnarray}
and obeys to the continuity equation:
\begin{eqnarray}
\frac{\partial\,\rho}{\partial\,t}+{\bfm\nabla}\cdot{\bfm j}=0 \ ,
\end{eqnarray}
which is the well-known Fokker-Planck equation. In the case of
interacting charged particles, Eq. (\ref{dob}) must be replaced by
the following NLSE:
\begin{eqnarray} \nonumber
&&i\,c\,\hbar\,D_0\,\psi=-\frac{\hbar^2}{2\,m}\,{\bfm D}^2
\,\psi+m\,\nu\,{\bfm\nabla}\left(\frac{{\bfm
J}_0}{\rho}\right)\,\psi\\
&&-2\alpha\frac{\hbar^2}{m}\left[\frac{\Delta\rho}{\rho}-
{1\over2}\left(\frac{{\bfm\nabla}\rho}{\rho}\right)^2\right]\psi+i\frac{\hbar}{2}
\nu\frac{\Delta\rho}{\rho}\psi \ ,\label{dob1}
\end{eqnarray}
whose associated nonlinear potential is given by:
\begin{eqnarray}
\nonumber U([\rho],\,[S],\,{\bfm A})\!\!&=&\!\!
\frac{\nu\,e}{2\,c}\,\left[\rho\,{\bfm\nabla}\,({\bfm\nabla}\,S-\!{\bfm
A})- {\bfm\nabla}\,\rho\cdot ({\bfm\nabla}\,S-\!{\bfm A})
\right]\\
&+&\alpha\,\frac{\hbar^2}{m}\,\frac{({\bfm\nabla}\,\rho)^2}{\rho}\
\ .\label{pot}
\end{eqnarray}
We note that the current ${\bfm J}_0$ is given by:
\begin{eqnarray}
\nonumber {\bfm
J}_0=-\frac{i\hbar}{2m}\!\left[\psi^\ast\!\left({\bfm\nabla}-\!\frac{ie}{\hbar
c}{\bfm A}\right)\psi-\psi\left({\bfm\nabla}+\!\frac{ie}{\hbar
c}{\bfm
A}\right)\!\psi^\ast\right] \ .\\
\end{eqnarray}
The current $\bfm J$ of the system (\ref{dob1}) assumes the form
${\bfm J}={\bfm J}_0-\nu\,{\bfm\nabla}\rho$ or explicitly:
\begin{eqnarray}
{\bfm J}=\frac{e}{m\,c}\,\rho\,({\bfm\nabla}S-{\bfm
A})-\nu\,{\bfm\nabla}\rho \ .
\end{eqnarray}
Performing the unitary transformation
$\psi\!\rightarrow\!\phi={\cal U}\psi$; $A_\mu\rightarrow A_\mu$
defined through Eqs. (\ref{trasf2})-(\ref{sig}), we obtain
\begin{eqnarray}
{\cal U}=\exp\left(-i\,\frac{m}{\hbar}\,\nu\,\log{\rho}\right) \
.\label{tr}
\end{eqnarray}
It is easy to verify that Eq. (\ref{dob1}) is transformed as:
\begin{eqnarray}
&&\nonumber i\,c\,\hbar\,D_0\,\phi=-\frac{\hbar^2}{2\,m}\,{\bfm
D}^2\,\phi\\
&&+\left(m\,\nu^2
-2\,\alpha\,\frac{\hbar^2}{m}\right)\,\left[\frac{\Delta\,\rho}{\rho}
-{1\over2}\,\left(\frac{{\bfm\nabla}\,\rho}{\rho}\right)^2\right]\,\phi
\ .\label{dob2}
\end{eqnarray}
We note that the transformation (\ref{tr}) is the same previously
introduced by Doebner and Goldin \cite{Doebner2,Doebner3}. The
problem concerning the elimination of the imaginary part in the
nonlinearity of Doebner-Goldin equation has been considered
previously in Ref. \cite{Doebner4}.

Equation (\ref{dob2}) can be obtain considering the gauge
transformation $\psi\rightarrow\psi;\,\,A_\mu\rightarrow\chi_\mu$
with:
\begin{eqnarray}
{\bfm\chi}&=&{\bfm
A}+\frac{m\,c}{e}\,\nu\,\frac{{\bfm\nabla}\,\rho}{\rho} \ ,\\
\chi_0&=&A_0+\frac{\nu}{c\,\rho}\,{\bfm\nabla}[({\bfm\nabla}\,S-{\bfm\chi})\,\rho]
\ .
\end{eqnarray}
We conclude the discussion on the gauged Doebner and Goldin
equation, by observing that Eq. (\ref{dob2}) can be written in the
form:
\begin{eqnarray}
i\,c\,\overline{\cdot}\hspace{-1.5mm}k\,D_0\,\phi=-\frac{{\overline{\cdot}\hspace{-1.5mm}k}^2}{2\,m}\,{\bfm
D}^2\,\phi \ ,\label{dob3}
\end{eqnarray}
where
$\overline{\cdot}\hspace{-1.5mm}k^2=\hbar^2\,(1+8\,\alpha)-4\,m^2\,\nu^2$.
The procedure used to transform Eq. (\ref{dob2}) into Eq.
(\ref{dob3}) is the same used in the case $A_\mu=0$, described in
Ref. \cite{Guerra}.

In conclusion we have shown here two different ways to reduce the
complex nonlinearity of a NLSE into a real one. In order to obtain
this, we can perform a nonlinear unitary transformation on the
matter field of Eq. (\ref{trasf1}) or, alternatively, by
performing a transformation on the gauge field
(\ref{trasfgauge1}). As a working example of the method here
discussed we have considered the gauged Doebner and Goldin
equation, describing a system of interacting charged particles.

\vfill\eject

\begin{thebibliography}{99}

\bibitem{Kaniadakis} G. Kaniadakis, and A.M. Scarfone: Rep. Math. Phys. (2000), in press.

\bibitem{Doebner1} H.-D. Doebner, and G.A. Goldin: Phys. Lett. A
\bfm{162} (1992), 397-401.

\bibitem{Doebner2} H.-D. Doebner, and G.A. Goldin: J. Phys. A: Math. Gen
\bfm{27} (1992), 1771-1780.

\bibitem{Doebner3} H.-D. Doebner, and G.A. Goldin: Phys. Rev. A \bfm{54}
(1996), 3764-3771.

\bibitem{Nettermann} P. Nettermann, and R. Zhdanov: J. Phys. A:
Math. Gen \bfm{29} (1996), 2869-2886.

\bibitem{Doebner4} H.-D. Doebner, G.A. Goldin, and P. Nettermann:
J. Math. Phys. \bfm{40} (1999), 49-63.

\bibitem{Noether} E. Noether: {\sl Invariante
Variationsprobleme}, Nachr. d. K\"onig. Gesellsch. d. Wiss. zu
G\"ottingeen, Math-phys. Klasse, 235, 1918; English translation:
M. A. Travel, Transport Theory and Statistica Physics 1(3), 183,
1971.

\bibitem{Olver} P. J. Olver: {\sl Applications of Lie Groups to
Differential Equations}, Springer, New York, 1986.

\bibitem{Jackson} J. D. Jackson: {\sl Classical Electrdynamics},
Wiley, New York, 1975.

\bibitem{Guerra} F. Guerra, and M. Pusterla: Lett. Nuovo Cimento
\bfm{34} (1982), 356.

\end{thebibliography}
\end{document}